\documentclass[twocolumn,showpacs,preprintnumbers,amsmath,amssymb, superscriptaddress]{revtex4}

\usepackage{graphicx}
\usepackage{dcolumn}
\usepackage{bm}

\begin{document}
\newcommand{\AlOx}{Al$_2$O$_3$ }

\preprint{Submitted to PRL}

\title{Enhanced magneto-transport at high bias in quasi-magnetic tunnel junctions with EuS spin-filter barriers}

\author{T. Nagahama}%
\affiliation{%
Francis Bitter Magnet Laboratory, Massachusetts Institute of Technology, Cambridge, MA 02139, USA
}%
\affiliation{%
NanoElectronics Research Institute, AIST, Tsukuba Central 2 Umezono 1-1 Tsukuba Ibaraki 305-8568, Japan
}%
\author{T. S. Santos}%
\author{J. S. Moodera}%
\affiliation{%
Francis Bitter Magnet Laboratory, Massachusetts Institute of Technology, Cambridge, MA 02139, USA
}%


\date{\today}

\begin{abstract}
In quasi-magnetic tunnel junctions (QMTJs) with a EuS spin filter tunnel barrier between Al and Co electrodes, we observed large magnetoresistance (MR). The bias dependence shows {\it an abrupt increase} of MR ratio in high bias voltage, which is contrary to conventional magnetic tunnel junctions (MTJs). This behavior can be understood as due to Fowler-Nordheim tunneling through the fully spin-polarized EuS conduction band. The I-V characteristics and bias dependence of MR calculated using tunneling theory shows excellent agreement with experiment. 
\end{abstract}

\maketitle

Magnetic tunnel junctions are commonly used in magnetoresistive random access memories and in read heads of hard disk drives.  For conventional MTJs, having two ferromagnetic electrodes separated by a non-magnetic insulator such as \AlOx or MgO \cite{mooderatmr, miyazaki, nagahamaMgOrev}, the tunnel magnetoresistance (TMR) consistently decreases with increasing applied voltage.  This bias dependence has been attributed to the excitation of magnons, phonons and band effects at higher bias \cite{mooderaiets}.  Consequently, the operation of a practical device utilizing a MTJ is limited to low bias for optimum TMR effect.  This Letter shows a novel way to overcome this limitation---a bias dependence behavior in which TMR {\it increases} with applied voltage, by utilizing a {\it ferromagnetic} tunnel barrier, EuS.

The ferromagnetic semiconductor EuS has the unique capability to filter electron spins, producing a nearly fully spin-polarized current \cite{sfreview}. This spin filter phenomenon arises from the large exchange splitting of the EuS conduction band, lowering the spin-up sublevel by 2$\Delta E_{ex}$=0.36eV from the spin-down sublevel \cite{Wachter}.   When EuS is used as a tunnel barrier, this exchange splitting produces a lower spin-up barrier height ($\Phi _\uparrow $) than spin-down barrier height ($\Phi _\downarrow $).  Because the spin-up (down) tunnel current is exponentially dependent on the spin-up (down) barrier height \cite{Brinkman},  a highly spin-polarized current is produced.  Whereas conventional ferromagnets (FMs) used as a source of polarized spins generally have a spin polarization P$<$50\%, the spin filter effect can reach P=100\% \cite{sfreview}.

The spin filter effect in a EuS tunnel barrier was first demonstrated by Esaki {\it et al.} with either Al or Au electrodes \cite{esaki}.  Later, field-emission experiments using EuS-coated tungsten tips showed high $P\sim 90\%$ resulting from spin filtering in EuS \cite{muller}.    Moodera {\it et al.} extensively studied the spin filter effect of ultrathin EuS tunnel barriers via the Meservey-Tedrow technique \cite{MTreview}, whereby P of the tunnel current was directly measured using a superconducting Al electrode as the spin detector, yielding P as high as 85\% for EuS just 3nm thick \cite{moodera, hao}.  The Meservey-Tedrow technique uses the superconducting quasi-particle density of states of Al in an applied field to directly determine P, and only probes within $\pm$2 meV of the Fermi level.  The spin-filter effect at high bias was still an unexplored phenomenon, up until this present work.

A versatile approach for utilizing the spin-filter effect is to incorporate the spin filter in a QMTJ device, thereby yielding a large MR.  In such a QMTJ the EuS tunnel barrier is sandwiched between a FM and a non-magnetic electrode.  Resistance of the junction $R_J$ depends on the relative alignment of the magnetization of the spin filter and the FM, resulting in higher resistance for antiparallel alignment ($R_{AP}$) and lower resistance for parallel alignment ($R_P$), where MR= $(R_{AP}-R_P)/R_P =\Delta R/R_P$.  LeClair {\it et al.} observed $> 100\%$ MR in QMTJs of Al/EuS/Gd \cite{leclair}.   However, the curve of $\Delta R/R_P$ versus applied field H was unstable, especially in the antiparallel state. This instability may be the result of exchange coupling between the EuS and Gd. In their work the measurement was performed in low bias, $< \pm$100 mV, whereas the behavior at higher bias is unknown. Noticeable spin filtering has been demonstrated more recently in QMTJs utilizing other spin filter materials, ${\rm NiFe_2O_4}$ \cite{luders} and ${\rm BiMnO_3}$ \cite{gajek}. In these studies with limited bias range, the MR ratio decreased monotonically as bias voltage increased, the same as in conventional MTJs.

The bias dependence of MR for a QMTJ with a spin filter barrier can be expected to show a novel behavior. Due to the spin-split conduction band of the spin filter, which we refer to here as a half conduction band (HCB), one should see an increase of MR for applied bias $V > \Phi _\uparrow$. This increase originates because for $V > \Phi _\uparrow$, the spin-up electrons have a preferred tunneling path through the spin-up HCB, which is not the case for spin-down electrons.  To verify this hypothesis, we investigated the tunneling characteristics at large V with a QMTJ consisting of a non-FM/EuS/insulator/FM, clearly showing a unique increase of MR as bias increased beyond a certain value.  We put forward a model to explain this behavior and our calculations based on this model are in agreement with the measurement. 

The QMTJ of the structure 5 nm Al/EuS(x)/\AlOx(y)/10 nm Co/CoO were prepared on glass substrates in a high vacuum deposition chamber with a base pressure of ${\rm 6x10^{-8}}$ Torr. The junctions were patterned {\it in situ} with shadow masks into a cross configuration with a junction area of 200 x 400 ${\rm \mu m^2}$.  The EuS barrier and the \AlOx layer were evaporated by electron-beam from EuS pellets and Sapphire pieces, respectively. The bottom Al electrode, EuS and \AlOx were deposited onto liquid nitrogen-cooled substrates, in order to form uniform, continuous films. The top 10 nm thick Co electrode was deposited at room temperature and then oxidized by exposure to oxygen plasma ($\sim$ 25 s) to form thin CoO, which acts as an exchange bias pinning layer. The characterization of tunnel junctions with EuS barriers is described in Ref. \cite{hao} in detail.  The current-voltage (I-V) characterization and MR measurement were carried out by a 4-terminal technique. The bias voltage was defined with respect to the top Co electrode.

Fig. 1(a) shows a MR curve for one of the junctions described above at 4.2 K with a bias of +1200 mV. A clear change in junction resistance due to the change in the relative magnetic configuration between EuS and Co is observed. The MR is shifted to positive magnetic field because the magnetization of Co is pinned by the CoO layer. Compared with the MR curve of LeClair et al. \cite{leclair}, the shape of the MR curve for this QMTJ is well-defined and reproducible for all QMTJs measured in this study.  This can be because the magnetizations of Co and EuS are well-separated by the \AlOx layer, preventing magnetic coupling between them.
	
\begin{figure}
\includegraphics[width=7cm]{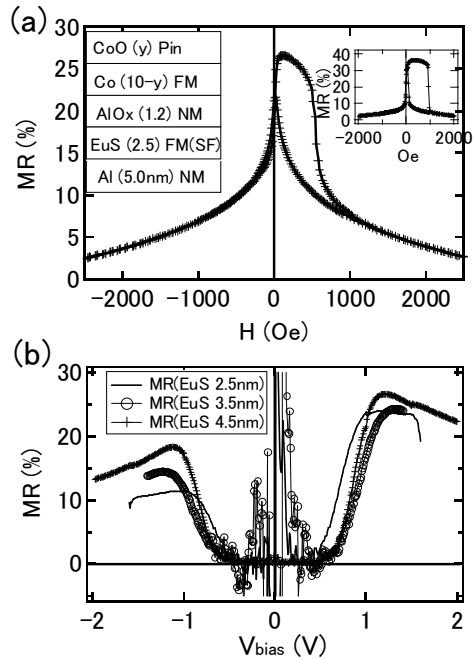}
\caption{\label{fig1} (a) Magnetoresistance of a 5 nm Al/2.5nm EuS/1.2 nm \AlOx/10 nm Co/CoO junction at 4.2 K with a bias of +1200 mV. The inset shows the MR curve for another junction showing the clearest MR. (b) Bias dependence of MR ratio for QMTJs with various thicknesses of EuS. In low bias voltage, the data is scattered due to the very large junction resistances.}
\end{figure}

To observe the bias dependence of MR ratio, we measured I-V curves at +200 Oe and +4000 Oe for antiparallel and parallel magnetization configuration of EuS and Co, respectively. Figure 1(b) shows bias dependences of MR ratios for 5 nm Al/EuS (x=2.5,3.5,4.5)/1.2 nm \AlOx/10 nm Co/CoO, having three thickness of EuS. The MR ratio increases dramatically at $\sim$800 mV for all three thickness of EuS. This is an unusual bias dependence because in conventional MTJs, the TMR ratio decreases monotonically with bias voltage \cite{mooderaiets}.  In low bias voltage, although the junction should show MR effect, it is unclear because the junction resistance was too high to measure, $>10^9$ Ohms. Therefore, we focused on the high bias region.

In order to understand the bias dependence of the MR ratio, we divided the data into two regions: $V_{bias}<\Phi_\uparrow$  and  $V_{bias}>\Phi_\uparrow$. In Fig. 2 the energy diagram for each region is schematically shown. In low bias voltage, $V_{bias}<\Phi_\uparrow$ (Fig. 2(a)), electrons must tunnel through both the EuS and \AlOx barriers (direct tunneling). Whereas in high bias voltage, $V_{bias}>\Phi_\uparrow$, a significant change occurs in the energy diagram (Fig. 2(b)). The spin-up conduction band is below the Fermi level (${\rm E_F}$) of the Al electrode, allowing only spin-up electrons to tunnel via the HCB, which has a greater tunneling probability than for direct tunneling. Therefore, P of tunnel current increases for $V_{bias}>\Phi_\uparrow$, causing an abrupt increase of MR ratio. At even higher bias, the spin-down conduction band also lowers below ${\rm E_F}$ of Al, resulting in a gradual reduction of MR ratio.

\begin{figure}
\includegraphics[width=7cm]{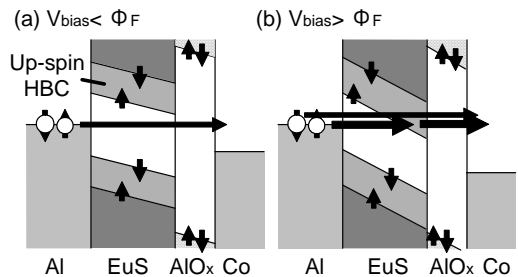}
\caption{\label{fig2}A schematic of the band diagram for the QMTJ: (a) $V_{bias} < \phi _{\uparrow }$. Electrons must tunnel through both the EuS and \AlOx barriers. (b) $V_{bias} > \phi _{\uparrow }$. Spin-up electrons tunnel via the HCB in EuS. Whereas spin-down electrons must still tunnel directly through both barriers.}
\end{figure}

In this model at high bias voltage (Fig.2 (b)), Fowler-Nordheim (FN) tunneling \cite{fn} takes place to the HCB. This is evident in the I-V characteristic of the junctions shown in Fig. 3. Although the I-V and dI/dV curves appear to be just exponential curves (see Fig. 3(a)), in a plot of d/dV(log(dI/dV)) versus V, large structures are apparent. Within $\pm$200 mV, the I-V characteristics show anomalous behavior, which originates from the zero bias anomaly---inelastic tunneling due to impurity states and magnons or phonons, which is common for MTJs \cite{wolf}. One can also see large peak structures at $\pm$0.8 V, which correspond to the voltage where the MR ratio increases (Fig.3 (b)). This peak indicates a transition from direct tunneling to FN tunneling.

\begin{figure}
\includegraphics[width=7cm]{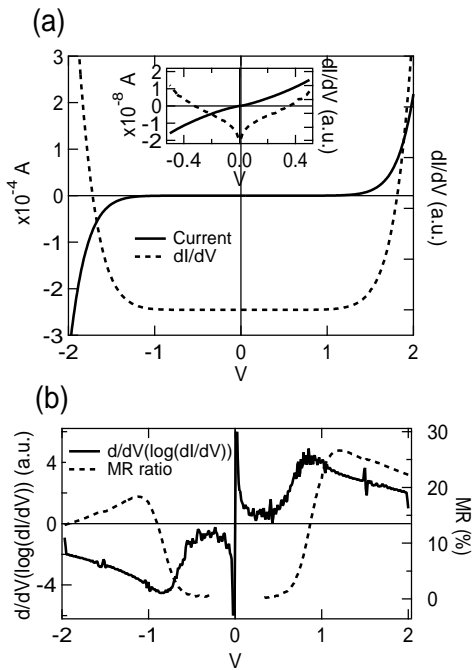}
\caption{\label{fig3}I-V characteristics of a 5 nm Al/4.5 nm EuS/1.0 nm \AlOx/10 nm Co/CoO junction. (a) The solid line is the I-V curve. The broken curve is the dI/dV curve. The inset is the plot within $\pm$ 500 mV. (b) The solid curve is d/dV(log(dI/dV)). The broken line is the MR ratio.}
\end{figure}

The physical meaning of d/dV(log(dI/dV)), which is equal to ${\rm (d^2I/dV^2)/(dI/dV)}$, is  explained as follows. To a first approximation, the differential conductance can be expressed as \cite{wiesendanger}
\begin{equation}
\frac{\text{dI}}{\text{dV}} \propto T(\text{E}) N_1(\text{E}) N_2(\text{E}),\ T(E)=\exp (f(\text{E}))
\end{equation}
where T(E) is tunneling probability and N(E) is  the density of states in the electrode. When we employ the WKB approximation  for T(E), it becomes an exponential function exp(f(E)). Since logN(E) is constant relative to log[exp(f(E))], d/dV(log(dI/dV)) is expressed as,
\begin{equation}
\frac{\rm d}{\text{dV}}\left[\log \left(\frac{\text{dI}}{\text{dV}}\right)\right]\cong \frac{{\rm d} f(\text{E})}{\text{dV}} .
\end{equation}
In other words, the physical meaning of d/dV(log(dI/dV)) is just the exponent of the tunneling probability T(E).

In order to interpret the peak in d/dV(log(dI/dV)), we calculated the tunnel probability T(E) by using the WKB approximation. Under the assumption of the trapezoidal barrier potential $V_{bias}$ (see Fig. 4(a) inset), i.e. $V(z)=\text{E}_F+\Phi _F-\frac{eV_{\text{bias}}}{d}z$, the tunneling probabilities for direct tunneling $T(E)_{DT}$ and FN tunneling $T(E)_{FN}$ are as follows.
\begin{multline}
T(\text{E})_{\text{DT}}=\exp\{\frac{-4 \sqrt{2m} d }{3 \hbar \text{eV}_{\text{bias}}}((E_F+\Phi_F-E_z)^{3/2}- \\
 (E_F+\Phi _F-E_z-\text{eV})^{3/2})\}
\end{multline}

\begin{equation}
T(\text{E})_{\text{FN}}=\exp \left\{\frac{-4 \sqrt{2m} d }{3 \hbar \text{eV}_{\text{bias}}} \left(E_F+\Phi_F-E_z\right)^{3/2}\right\}
\end{equation}
where \textit{m} is the electron mass, $\Phi _F$ is the barrier height from $\rm{E_F}$, and \textit{$E_z$} is energy component of electrons normal to the barrier surface. 

Fig. 4(a) shows ${\rm d/dV(log T(E_F))}$ as a function of ${\rm V_{bias}}$, assuming a barrier height ${\rm \Phi _F}$ of 0.8 eV. The value is higher than that reported by Moodera {\it et al.} \cite{moodera}; it is estimated due to existence of a thin \AlOx layer between EuS and Co. The plot agrees very well with the experimental data shown in Fig. 3(b), except around zero bias voltage where the experimental data shows large value due to aforementioned reasons. We carried out a numerical calculation of the tunnel current based on the WKB approximation and free electron model \cite{duke, floyd},  which also showed good agreement with experiment \cite{nagahama} (not shown). The results of these calculations identify the peaks in Fig. 4 to occur at the barrier height. Above the peak voltage FN tunneling, i.e. tunneling through the HCB, dominates the transport. This strongly supports our model of increasing MR due to tunneling via the HCB.

\begin{figure}
\includegraphics[width=7cm]{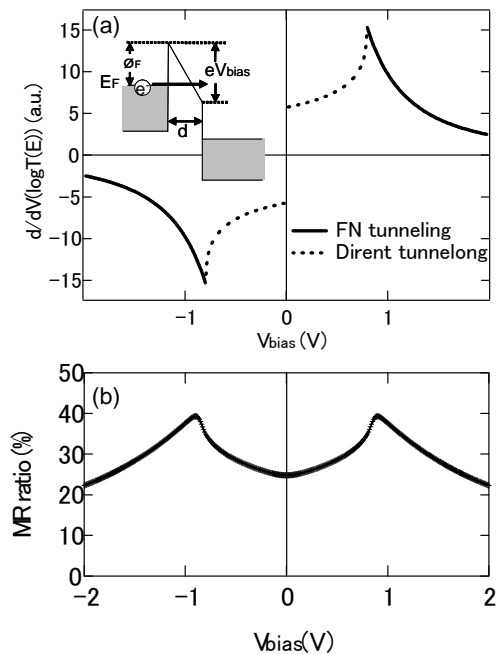}
\caption{\label{fig4} (a) Calculated result of d/dV(logT(E)). The inset is the energy diagram used in the calculation. (b) Calculated MR ratio as a function of bias voltage.}
\end{figure}

Fig. 4(b) is a simulation of the bias dependence of the MR using Julliere's formula \cite{jullier} with spin polarization of the tunnel current due to spin-filtering in EuS (${\rm P_{tc}}$), and spin polarization of Co (${\rm P_{Co}}$).  ${\rm P_{tc}}$ is estimated from a tunnel current calculation for spin-up and spin-down current. The parameters for the simulation are as follows: $\Phi _\uparrow $=0.8 V, $\Phi _\downarrow $=0.85 V and ${\rm P_{Co}}$ = 0.2, which is smaller than the reported value for Co in low bias. ${\rm P_{Co}}$ should vary with bias, whereas its bias dependence is unknown. Hence we fixed the value at ${\rm P_{Co}}$=0.2 for the first approximation.  In the plot for this calculation, Fig. 4(b), MR rises around 0.8 V (which is $\Phi _\uparrow $) and decreases gradually at higher voltages, which is same behavior as in the experimental data of Fig. 2. Saffarzadeh has done similar calculations for a more ideal situation, and the same tendency was observed \cite{saffarz}.  The exchange splitting of 0.05 V used in the simulation is smaller than the value in Ref. \cite{moodera}. This low value of the splitting can be justified because the inelastic scattering or spin flip scattering due to magnon or magnetic impurity etc. are not taken into account in this simple model. It is well known that such scattering reduces the MR ratio significantly in high bias voltage \cite{zhang, mooderaiets}. It should be noted that the peak position of d/dV(log(dI/dV)) is $\sim$0.8 eV in the Fig. 3(b) but a part of that bias is applied across the \AlOx barrier. Even if that is the case, the variation of the barrier height does not change the result of the calculation qualitatively. Because the junction structure is asymmetric, one could expect the I-V curve to be asymmetric, whereas the I-V curve in Fig. 3 is symmetric.  This could be due to the presence of defect states in the barrier or interfaces \cite{gibson}. In low bias voltage some discrepancies exist between the simulation and the experiment (Fig. 4(b)).  However, the measurement was not stable in this region because of the high junction resistance. More careful measurements in the low bias region will be done in the future.

In conclusion, we combined the spin filter phenomenon with the fully spin-polarized half conduction band of EuS to obtain a unique MR behavior, using quasi-magnetic tunnel junctions with EuS spin-filter barriers. The junction shows abrupt increase of MR at some intermediate bias voltage of ${\rm \pm}$ 0.8V. This has been attributed successfully to the tunneling via fully spin-polarized half conduction band in EuS, and confirmed by a model calculation of tunnel current including Fowler-Nordheim tunneling. Although one can improve MR to a great extent by carefully tailoring the growth of a spin filter barrier, our current results show the possibility to reach high spin polarization and MR at large biases. This is ideally suitable for spin-based devices \cite{molnar, geballe}, for example for effective large spin polarized carrier injection into a semiconductor.
\begin{acknowledgments}
This work was supported by NSF and ONR grants.  Partial support came from the KIST-MIT project funds.  T.N.'s stay at MIT was supported by research funds from AIST, Tsukuba, Japan.
\end{acknowledgments}

\bibliography{EuSPRL_2}

\end{document}